%Paper: hep-th/9507121
%From: Edward Witten <witten@sns.ias.edu>
%Date: Sun, 23 Jul 1995 10:01:30 EDT

\input harvmac
\newcount\figno
\figno=0
\def\fig#1#2#3{
\par\begingroup\parindent=0pt\leftskip=1cm\rightskip=1cm\parindent=0pt
\baselineskip=11pt
\global\advance\figno by 1
\midinsert
\epsfxsize=#3
\centerline{\epsfbox{#2}}
\vskip 12pt
{\bf Fig. \the\figno:} #1\par
\endinsert\endgroup\par
}
\def\figlabel#1{\xdef#1{\the\figno}}
\def\encadremath#1{\vbox{\hrule\hbox{\vrule\kern8pt\vbox{\kern8pt
\hbox{$\displaystyle #1$}\kern8pt}
\kern8pt\vrule}\hrule}}

\overfullrule=0pt

%macros
%
\def\tilde{\widetilde}

\font\zfont = cmss10 %scaled \magstep1

\def\bigone{\hbox{1\kern -.23em {\rm l}}}
\def\ZZ{\hbox{\zfont Z\kern-.4emZ}}

\Title{hep-th/9507121, IASSNS-HEP-95-63}
{\vbox{\centerline{Some Comments On String Dynamics}}}\foot{To appear
in the proceedings of {\it Strings `95}, USC, March      1995}
\smallskip
\centerline{Edward Witten}
\smallskip
\centerline{\it School of Natural Sciences, Institute for Advanced Study}
\centerline{\it Olden Lane, Princeton, NJ 08540, USA}\bigskip

\medskip

\noindent
%write abstract here
Three subjects are considered here:
a self-dual non-critical string that appears
in  Type IIB superstring theory at points in ${\rm K3}$ moduli space where
the Type IIA theory has extended gauge symmetry; a conformal
field theory singularity at such points which may signal
quantum effects that persist  even at
weak coupling; and the rich dynamics of the real world under
compactification, which may be relevant to some attempts to
explain the vanishing of the cosmological constant.

\Date{July, 1995}
%text of paper
\def\R{{\bf R}}
\def\S{{\bf S}}
\def\T{{\bf T}}
\def\K3{{\rm K3}}
My lecture at {\it Strings `95} focussed on determining the strong
coupling behavior of various string theories in various dimensions.
Among the main points were the following: $U$-duality of Type II
superstrings requires that the strong coupling limit of the
Type IIA superstring in ten dimensions is eleven-dimensional supergravity
(on ${\bf R}^{10}\times{\bf S}^1$); one can make sense
of heterotic string dynamics in five, six, and seven dimensions
and deduce $S$-duality in four dimensions if one assumes that
the heterotic string on ${\bf R}^6\times {\bf T}^4$ is equivalent
to the Type IIA theory on ${\bf R}^6\times {\rm K3}$.  The detailed
arguments have appeared elsewhere \ref\witten{E. Witten,
``String Theory Dynamics In  Various Dimensions,'' hepth-9503124,
to appear in Nucl. Phys. B.} and will not be repeated here.
Instead I will try to clarify a few related issues, in some cases
involving questions that were asked at the meeting.

The issues I will discuss in sections one and two
involve mainly the extended gauge symmetry of the Type IIA
superstring on ${\bf R}^6\times {\rm K3}$ at certain points in moduli
space.  In    section one, I analyze how the Type IIB theory behaves
when Type IIA has extended gauge symmetry, and in section two,
I discuss the nature of the singularity that occurs in conformal
field theory at these points.  In section three,    I consider
instead some issues involving the behavior of the real world
under dimensional reduction; these issues may be relevant
to the vanishing of the cosmological constant.

\newsec{The Type IIB Theory On ${\bf R}^6\times \K3$}

The best-established string-string duality is the equivalence
between the heterotic string on $\R^6\times \T^4$ and the Type IIA
string on $\R^6\times\K3$.  According to this equivalence, the   Type IIA
model on $\R^6\times \K3$ gets extended non-abelian gauge symmetry
at certain points in $\K3$ moduli space.  Our first question is to
determine  how the Type IIB theory -- likewise compactified on
$\R^6\times \K3$ -- behaves at the
 points in moduli space at which
the Type IIA theory  develops enhanced gauge symmetry.

\nref\aspinwall{P. Aspinwall and D. Morrison, ``$U$-Duality And
Integral Structure,'' hepth-9505025.}
It is certainly not the case that the Type IIB theory develops enhanced
gauge symmetry at those points.  In fact, the  Type IIB
theory on ${\bf R}^6\times\K3$ has a chiral supersymmetry that simply
does not admit gauge multiplets of any kind, abelian or non-abelian.
It is very hard for the Type IIB theory to get extra massless
particles at special points in $\K3$ moduli space, and these are not
needed to account for singularities in the Zamolodchidov metric.
  The moduli space of vacua of the Type IIB theory
on ${\bf R}^6\times \K3$ is apparently \refs{\witten,\aspinwall}
the locally homogeneous space $SO(21,5;{\bf Z})\backslash
 SO(21,5;\R)/(SO(21)
\times SO(5))$.  The singularities of this space are orbifold
singularities, and instead of looking for a description of these
in terms of extra massless particles, we can simply interpret
them as a sign of restoration of a discrete local gauge symmetry.

However, such discrete symmetry restoration cannot be the whole
story of what happens to the Type IIB theory at special points
in $\K3$ moduli space.  This becomes clear if one makes a further
compactification to $\R^5\times \S^1\times \K3$.  Once this is done,
the Type IIB theory becomes equivalent to the Type IIA theory,
which does get extra massless particles at certain points in
$\K3$ moduli space.  The Type IIB theory has to do something peculiar
such that one does not get extra massless particles on $\R^6\times \K3$,
but one does get extra massless particles on $\R^5\times \S^1\times\K3$,
for any radius of the circle.

Let us write down the precise comparison of the Type IIA and Type IIB
theories in this situation. We will be a little more general than
the six-dimensional case.  First, if a $d$-dimensional string
theory is compactified to $d-1$ dimensions on a circle of circumference
$R$, then the relation between the $d$ and $d-1$-dimensional string
coupling constants is
\eqn\relco{{1\over \lambda_{d-1}^2}={R\over\lambda_d^2}.}
For Type IIA and Type IIB theories compactified from $d$ to $d-1$
dimensions on a circle to be equivalent, they must have the
same $\lambda_{d-1}$, so the relation among couplings in $d$ dimensions
is
\eqn\melco{{R_A\over \lambda_{d,A}^2}={R_B\over \lambda_{d,B}^2}.}
Here $R_{A}$ and $R_B$ are the circumference of the circle as
measured in the Type IIA and Type IIB theories, and similarly
$\lambda_{d,A}$ and $\lambda_{d,B}$ are the respective string couplings.
Bearing in mind also the $T$-duality relation $R_A=1/R_B$, we can
write \melco\ as
\eqn\helco{{1\over \lambda_{d,A}}={R_B\over \lambda_{d,B}},}
a relation that of course also holds if $A$ and $B$ are exchanged.
We will henceforth write the $d$-dimensional couplings as simply
$\lambda_A$ and $\lambda_B$.

Now, suppose that in the $\K3$ moduli space, one is a distance $\epsilon$
from a point at which the Type IIA theory gets an enhanced gauge
symmetry.  Then the Type IIA theory on $\R^6\times \K3$
has $W$ bosons with
mass a constant times $\epsilon/\lambda_A$; this mass is unchanged
in  compactification to $\R^5\times \S^1\times \K3$.
(The $W$ mass  is exactly independent of $R_A$, not
just approximately so for large $R_A$, because the $W$ boson
is in a BPS-saturated  ``small'' supermultiplet.)
According to \helco, the mass  of the $W$ meson in the Type IIB
theory   on $\R^5\times \S^1\times\K3$ is then
\eqn\hudn{M_W={\epsilon R_B\over\lambda_B}.}

What are we to make of \hudn?  What sort of state has a mass proportional
to $R_B$?  The answer to this question, clearly, is that this is the
mass of a string wrapping around the circle of circumference $R_B$.
So we can interpret \hudn\ to mean that the Type IIB theory on
$\R^6\times \K3$ has some kind of cosmic string with a string
tension
\eqn\udn{T={\epsilon\over\lambda_B}.}
After compactification on a circle, the $W$ boson then arises
as a particular winding state of this string.

The string whose tension is given    in  \udn\ is certainly
not the fundamental Type IIB superstring.  Rather, we must
apparently begin with the self-dual super-three-brane solution of the
Type IIB theory in ten dimensions \ref\duff{M. J. Duff
and J. X. Lu, ``The Self-Dual Type IIB Superthreebrane,''
Phys. Lett. {\bf B273} (1991) 409.}, whose tension is of order
$1/\lambda_B$.  As $\epsilon$ goes to zero, a two-sphere $\S$ in the
$\K3$ collapses, having an area proportional to $\epsilon$
\witten.
$\S$ is self-dual, in the sense that its Poincar\'e dual cohomology
class is self-dual.
As in Strominger's discussion of the conifold singularity in four
dimensions
\ref\strominger{A. Strominger, ``Massless Black Holes And Conifolds
In String Theory,'' hepth-9504090.},  when one compactifies
below  ten dimensions, one can get a $p$-brane for $p<3$ by wrapping
the ten-dimensional super-three-brane around a cycle of dimension
$3-p$.  In particular, upon $\K3$ compactification, one can
wrap the three-brane around $\S$ to get a string in six dimensions.
The tension of this string will be $1/\lambda_B$ (the tension
of the three-brane) times $\epsilon$ (the area of $\S$), in agreement
with \udn.  Since $\S$ is self-dual, the string we get in six
dimensions is likewise self-dual (that is, the three-form
$H=dB$ that the string produces is self-dual).
It is thus similar to the six-dimensional self-dual string
described in \ref\duffo{M. J. Duff and J. X. Lu, ``Black And
Super $p$-Branes In Diverse Dimensions,'' Nucl. Phys. {\bf B416}
(1994) 301.}.

This self-dual string is a non-critical string in six dimensions;
its tension \udn\ can be vastly below the string and Planck scales.
For very small $\epsilon$, one should interpret this as a string
that is far too light to influence gravity and which simply propagates
in six-dimensional Minkowski space.  There is such a
string theory for each possible type of isolated singularity
($A$, $D$, or $E$) of the $\K3$.  (The formulation in the last
paragraph with a single collapsing two-sphere was strictly
appropriate only for $A_1$.)
Obviously, these non-critical six-dimensional strings are quite
different from anything we really understand presently.
The fact that these objects have not been discovered in traditional
constructions of string theories actually follows from the fact
that they are self-dual, so that (as in Dirac quantization of electric
and magnetic charge) the string coupling is necessarily of order one.

A weakly coupled string theory with string tension $T$ has long-lived
excitations with     masses proportional to $\sqrt T$.  If this
formula can be used in the present case - which is not  entirely
clear - then the Type IIB theory near the special $\K3$ points
has long-lived
non-perturbative ``string'' states with masses in string units
proportional
to $\sqrt{\epsilon/\lambda_B}$.  In Einstein units, these states
have masses of order $\sqrt \epsilon$.
In heterotic string units, the mass is of order
$\sqrt{\epsilon/\lambda_h}$ (times the masses of elementary string
states), with $\lambda_h$ the heterotic string coupling constant.

\subsec{Reduction To Four Dimensions}

Now, let us recall that, although the Type IIB theory
on $\R^6\times \K3$ does not have any gauge fields, it does have
a plethora of two-forms (twenty-one with self-dual field strength
and five with anti-self-dual field strength).  One of them - say $B$ -
arises by writing the four-form $C$ of the ten-dimensional
Type IIB theory as $C= B\wedge G$, where  $G$ is a self-dual
harmonic two-form on $\K3$ supported (for small $\epsilon$) very
near $\S$, and $B$ is a two-form on $\R^6$.  $B$ is self-dual
(that is, it has a self-dual field strength) because $C$ and $G  $
are self-dual.

If we compactify from $\R^6$ to $\R^5\times \S^1$, then the
components $B_{i6}$ ($i=1\dots 5$)
of $B$ become a gauge field $A_i$ in five
dimensions.  One might think that one would also get a five-dimensional
two-form from $B_{ij}$, but in five-dimensions a two-form is dual
to a one-form, and self-duality of $B$ in six dimensions becomes
in five dimensions the statement that $B_{ij}$ is dual to $A_i$.  Thus
the independent degrees of freedom are all in $A_i$.
The string winding states discussed above carry the electric
charge that is coupled to $A_i$.

The  further compactification to four dimensions, replacing
$\R^6$ by $\R^4\times \T^2$, has been discussed at the
field theory level  in \ref\verlinde{E. Verlinde, ``Global Aspects
Of Electric-Magnetic Duality,'' hepth-9506011.}.
The self-dual two-form $B$ in
six dimensions again gives rise in four dimensions to only {\it one}
$U(1)$ gauge field - as $A_i=B_{i6}$ and $\tilde A_i=B_{i5}$
are dual.

Going back to string theory, it follows that the two types
of  winding states of the non-critical string
- strings wrapping around the first
or second circle in $\T^2=\S^1\times \S^1$  - carry electric and
magnetic charge for this one $U(1)$ gauge field.
The coupling parameter
$\tau$ of the   four-dimensional $U(1)$ theory is simply
the $\tau$ of the    $\T^2$.
The four-dimensional theory has manifest $S$-duality
coming from the diffeomorphisms of the $\T^2$.
(If we bear in mind that $SL(2,{\bf Z})_U$ of Type IIB
is $SL(2,{\bf Z})_T$ of Type IIA, this is equivalent to the fact
\witten\ that string-string duality transforms $S$-duality of the
heterotic string into $T$-duality of the Type IIA string.)

What makes this interesting is that it gives a manifestly $S$-dual
formulation of $N=4$ supersymmetric Yang-Mills theory.
In fact, for very small $\epsilon$, the  $W$ bosons and monopoles
(which come from string winding states and have masses of order
$\epsilon$) are much lighter than other string excitations (which
as we noted above generically have masses of order $\epsilon^{1/2}$).
Thus, in this limit, the manifestly $S$-dual theory of the
self-dual string on $\R^4\times \T^2$ should go over to $N=4$
supersymmetric Yang-Mills theory on $\R^4$.

This may well be the proper setting for understanding $S$-duality
of the $N=4$ gauge theory.  Thus, if one asks,  ``How can
the $S$-duality of $N=4$ Yang-Mills theory be made obvious?''  one
answer is that this can be  done by embedding $N=4$ supersymmetric
Yang-Mills theory in the heterotic string and then mapping to
a Type IIA theory by using string-string duality.  The weakness of
 this answer is that it    embeds  the     gauge theory
in a problem     with many other features - such as gravity - that
may not be material.  One would like to ``flow to the infrared,''
eliminating as many degrees of freedom as possible, and obtaining
the minimal theory in which the $S$-duality is still manifest.
The self-dual string in six dimensions may be the answer to this
question.

The self-dual string in six dimensions does not look easier  than
the Type IIB model that we started with; certainly we understand it less.
Nevertheless, it might be the right structure for understanding
the four-dimensional field theory.  The situation would be
somewhat similar to the study of critical phenomena.  In that subject,
one can start with an elementary,
 manifestly well-defined system such as a lattice
Ising model.  In seeking to describe the critical behavior, the right
object  to introduce turns out to be a continuum quantum field theory
even though this is superficially far less elementary (existence
is far less obvious, for instance) and superficially there are far
more degrees of freedom.  The field theory is the right object for
critical phenomena because it contains all the universal information
(about the critical point) and nothing else.
The more elementary-looking Ising model has the field theory as a
difficult-to-extract limit; the additional information it
contains is extraneous.  The self-dual string
may similarly  be the minimal manifestly $S$-dual extension of the
$N=4$ super Yang-Mills theory.

\subsec{Non-Local Critical Points In Four Dimensions}

Likewise, natural answers to other questions about gauge theory
dynamics may involve non-critical strings of one kind or another.
For instance, there appears to be \ref\argyres{P. Argyres and P. Douglas,
``New Phenomena In $SU(3)$ Supersymmetric Gauge Theory,''
hepth-9505062..}
an $N=2$ superconformal    critical point in four dimensions
with massless electrons and  monopoles alike.  A natural understanding
of this critical point may be  difficult to achieve in field theory
-- where it is hard to put electrons and monopoles on the same footing.
Perhaps one should seek  a natural description by some sort of
non-critical string theory.

Certainly critical string theory gives a natural framework
for describing generalizations of the critical point considered
in \argyres.  That critical
point, first of all, can be embedded in string
theory by simply considering a Calabi-Yau manifold with a singularity
that looks like
\eqn\hijji{t^2+w^2+y^2+x^3=\epsilon.}
This manifold contains \ref\milnor{J. Milnor, {\it Singular Points Of
Complex Hypersurfaces} (Princeton University Press, 1968).}
two $\S^3$'s that collapse as $\epsilon\to
0$.  These two     $\S^3$'s have a non-zero intersection number,
with the result that the charged black holes that arise as $\epsilon\to 0$
are respectively electrically and magnetically charged with respect
to the {\it same} $U(1)$.  In fact, the description of the critical
point in \argyres\ involves essentially the family of complex curves
\eqn\ijji{y^2+x^3=\epsilon.}
In this case, a pair of $\S^1$'s with a non-zero intersection
number collapse as $\epsilon\to 0$.  Obviously, \hijji\ is obtained
from \ijji\ by adding new variables that appear quadratically,
a standard operation that preserves many aspects of the singularity.

An $SU(N)$ generalization of this critical point that was
explained briefly
in \argyres\ involves
the family of curves $y^2+x^N=\epsilon$ and could be imitated
by a Calabi-Yau singularity $w^2+z^2+y^2+x^N=\epsilon$.  More generally,
the $N=2$ $SU(N)$ gauge theory with a massive adjoint hypermultiplet
can realize an arbitrary singularity of the form $F(x,y)=\epsilon$
\ref\donagi{R. Donagi and E. Witten, to appear.}, corresponding to a
Calabi-Yau singularity
\eqn\ikkl{t^2+w^2+F(x,y)=\epsilon.}

{}From the Calabi-Yau point of view, we can write many more objects,
such as a general hypersurface singularity $F(t,w,x,y)=0$.  To
restrict oneself to singularities that are at a finite distance in
the Zamolodchikov metric, one should consider what are called the
canonical singularities (reviewed in \ref\reid{M. Reid,
``Young Person's Guide To Canonical Singularities,'' {\it Algebraic
Geometry: Bowdoin, 1985}, Proc. Symp.  Pure Math. {\bf 46} (AMS, 1987).})
of which \ikkl\ is an example.  As there are many other canonical
singularities, it may turn out that the natural classification
and description of such non-local fixed points involves the
canonical singularities and the string theory dynamics they produce.

\newsec{The Singularity Of The Conformal Field Theory}

At the time of {\it Strings `95}, two points about the extended
gauge symmetry of the Type IIA superstring on ${\bf R}^6\times {\rm K3}$
were particularly puzzling:

(1) Although it was clear that the extended gauge symmetry occurred
only when one or more two-spheres collapse to zero area, it was
not clear why such collapse would lead to the appearance of
extra massless gauge bosons.

(2) More generally, it seemed that the collapse of a two-sphere could
lead to an interesting novelty in string theory
 only if there is some sort of breakdown
of the conformal field theory.  The example of orbifolds, which certainly
contain collapsed two-spheres (which are restored to non-zero area if one
blows up the orbifold singularities by adding suitable twist fields
to the world-sheet Lagrangian) seemed to show that there was absolutely
no singularity in the conformal field theory when a two-sphere collapses.

The first point was soon settled by Strominger \strominger:
a two-brane wrapped around a two-sphere goes
to zero mass when the two-sphere collapses to zero area.  (Strominger
discussed mainly compactification on Calabi-Yau threefolds,
 but the application to
$ \K3$ is evident.)

The second point was settled more recently by Aspinwall
\ref\newaspinwall{P. Aspinwall, ``Enhanced Gauge Symmetries
And $\K3$ Surfaces,'' hepth-9507012.}
who showed that extended gauge symmetry arises only when
there is a collapsed two-sphere {\it and in addition a certain
world-sheet theta angle vanishes}.  Orbifolds, that is $\K3$'s that
are of the form $\T^4/\Gamma$ with $\Gamma$ a finite group,
contain collapsed two-spheres, but the relevant theta angles are
non-zero.

In fact, associated with the
${\bf S}^2$ are four real parameters: the area, the theta angle,
and two parameters associated with the complex structure.  Aspinwall's
 claim is that all four
parameters must vanish to  get extended gauge symmetry.

Since orbifolds no longer serve as a counterexample, the likelihood
now arises that the $\K3$ conformal field theory is singular
at the points at which extended gauge symmetry appears.  That
is the question that I wish to address in the present section.
I will analyze the question by a mean field theory approach,
and suggest an answer that seems natural.  First we will look
at a problem -- which proves to be analogous -- of an instanton
shrinking to zero size; then we will consider the $\K3$ case; and finally
we will discuss conifold singularities of threefolds
in a similar spirit.

\subsec{Instanton Shrinking To Zero Size}

In \ref\ewitten{E. Witten, ``Sigma Models And The ADHM Construction
Of Instantons,'' J. Geom. Phys. {\bf 15} (1995) 215.},
I described a mean field approach to sigma models that are related
to Yang-Mills instantons.  This was achieved by constructing
two-dimensional linear sigma models with $(0,4)$ world-sheet
supersymmetry, which appear to flow in the infrared to conformal
field theories related to Yang-Mills instantons on ${\bf R}^4$.

I will not here recall the full details of the construction.  Suffice
it to note that the bosons are four massless fields $X^{BY}$, and
 additional fields $\phi^{B'Y'}$ that are generically massive (each of the
four types of index $B,B',Y,Y'$ is acted on by a different symmetry
group); inclusion of the massive fields
makes it possible to write a simple polynomial Lagrangian
that leads (after integrating them out) to very complicated
Yang-Mills instantons.  In the one-instanton sector, the description
is particularly simple; there are four $\phi$'s, and the potential
energy is
\eqn\potenergy{V={1\over 8}\left(X^2+\rho^2\right)\phi^2}
with $\rho$ the instanton size.  For $\rho$ large (compared
to the string scale), $\phi$ is everywhere very heavy, and after
integrating it out one gets something very much like
an ordinary Yang-Mills instanton, embedded in string theory.
For $\rho$ of order  the string scale, the stringy corrections
to the instanton may be large.  The point on which we wish to focus
here is the behavior of the conformal field theory when $\rho $ goes
to zero.  If we take \potenergy\ literally, we appear to learn
that the ``target space,'' obtained by setting $V=0$, acquires
a second branch precisely at $\rho=0$.  Apart from the usual
space-time $M $ with  $X$ unrestricted and $\phi=0$, we    get a second
world $M'$ with  $\phi$ unrestricted and $X=0$.   The linear
sigma model at $\rho=0$ in fact has a symmetry that exchanges $X$ and
$\phi$.

\nref\oldstrom{A. Strominger, ``Heterotic Solitons,'' Nucl. Phys.
{\bf B343} (1990) 167.}
\nref\harvey{C. G. Callan, Jr., J. A. Harvey, and A. Strominger,
``World-Sheet Approach To Heterotic Instantons And Solitons,''
Nucl. Phys. {\bf B359} (1991) 611.}
Before accepting the strange idea that when an instanton reaches
zero size, a second branch in space-time appears, let us compare
to another approach to the problem, in which one simply solves the
space-time equation for the instanton including terms of order
$\alpha'$ \refs{\oldstrom ,\harvey}.
In this approximation, the metric on a space-time that contains
an instanton of scale parameter $\rho$ centered on the origin turns
out to be
\eqn\guggle{ds^2=(dX)^2\cdot \left(e^{2\phi_0}
+8\alpha'{X^2+2\rho^2\over (X^2+\rho^2)^2}\right)}
(with $\phi_0$ the value of the dilaton at infinity).
The picture is quite different from what one seems to get from mean
field theory.  As $\rho$ goes to zero, instead of a second branch
appearing, the space-time develops a long tube, with the result
that at $\rho=0$, $X=0$ is infinitely far away.

It is true that \guggle\ is based only on solving the low energy
equations to lowest order in $\alpha'$.  However, one can show
\harvey\ that at $\rho=0$, the long tube that arises near $X=0$
corresponds to an exact soluble conformal field theory (a WZW model
times a free field with a linear dilaton), and this gives credence to the
idea that the structure seen in \guggle\ is essentially correct.

On the other hand, there is the following problem in the
``two-branch'' scenario that mean field theory seems to suggest.
The global $(0,4)$ supersymmetry algebra admits an $SU(2)\times SU(2)$
group of $R$ symmetries.  To extend the global $(0,4)$
algebra to a superconformal algebra, one of the two $SU(2)$'s
is included in the algebra and so is generated, in particular,
by purely right-moving currents.  (The second $SU(2)$ is not
part of the superconformal algebra, but might be realized as a symmetry
group acting by outer automorphisms on the algebra; if so the conserved
current generating this symmetry can have both left and right-moving
pieces.)

Now, the linear sigma model of the instanton has at $\rho=0$
the full $SU(2)\times
SU(2)$ symmetry, called $F\times F'$ in \ewitten.
If this model flows in the infrared to a $(0,4)$ superconformal
theory, is it $F$ or $F'$ that appears in the superconformal algebra?
The basic fact here is that $F$ acts by rotations of $X$ but
acts  trivially on $\phi$, and $F '$ rotates $\phi$ but acts trivially
on $X$.  The currents generating the $F$ action on $X$ are
$X^{AY}\partial_\alpha X^B{}_Y$ and have both left and right-moving
parts which are not separately conserved even if $X$ is treated as
a free field (which is valid for $X$ large enough); the currents
generating the $F'$ action on $\phi$ are similar.

Therefore, in any superconformal
description that contains $X$, $F$ cannot appear
in the superconformal algebra, and in any description that contains
$\phi$, $F'$ cannot appear in the superconformal algebra.
If this theory flows to a superconformal field theory in the infrared,
then (short of more exotic possibilities in which the pertinent symmetries
cannot be seen in the linear sigma model at all)
the symmetry between $X$ and $\phi$ must be  ``spontaneously broken'':
there must be two different superconformal limits, one living on the
$X$ branch with $F'$ in the       algebra, and one living on the $\phi$
branch with $F$ in the algebra.

Given that in the linear sigma model the distance between the $X$ and
$\phi$ branches appears to be finite (they even meet at $X=\phi=0$)
how is this possible?  It must be that as one flows to the infrared,
the distance grows from any given point on either branch to the point
$X=\phi=0$ where they meet; in the limit of the conformal field
theory, this distance must become infinite and the two branches
separate.  What makes this plausible is that near $X=\phi=0$, the
classical linear
sigma model does not give a good approximation to the metric on
the target space; loop diagrams are proportional to negative powers
of the mass, that is, to powers of  $1/X^2$ or $1/\phi^2$.

Thus, we have recovered, or at least rationalized, the qualitative
structure of \guggle\ from the linear sigma model.  To avoid a
contradiction with the properties of $F$ and $F'$, the two branches
must be decoupled at $\rho=0$, and this most reasonably happens
by $X=0$ being infinitely far away (from finite points on the $X$ space)
when $\rho=0$, as we see in \guggle.

\subsec{Singularities Of $\K3$'s}

Now I wish to describe a similar mean field theory by which
we can study orbifold singularities of $\K3$.  For simplicity,
we will discuss only the ${\bf Z}_2$ orbifold singularity,
so we will analyze simply the $(4,4)$ superconformal field theory
with target space $\R^4/{\bf Z}_2$.
\foot{
The general $A-D-E$ case can be discussed using a linear sigma
model constructed  via Kronheimer's
description of the $A-D-E$ singularities as hyper-Kahler quotients
\ref\kronheimer{P. Kronheimer, ``The Construction Of $ALE$ Spaces
As Hyper-Kahler Quotients,'' J. Diff. Geom. (1989) 665.}.
Kronheimer's construction
specializes for $A_1$ to the description we will give below, which
also entered in
 \ref\wstwo{N. Seiberg and E. Witten, ``Monopoles, Duality, and Chiral
Symmetry Breaking In $N=2$ Supersymmetric QCD.''}.}
In the twisted sector of this
orbifold, there are four moduli (three of them involving the classical
blow-up and deformation of the  singularity and one the world-sheet
theta angle).  Since it is difficult to add twist fields to the
Lagrangian with finite coefficients (as one must do, according
to \newaspinwall, to reach the point relevant to extended gauge
symmetry), we will study instead a $(4,4)$ linear sigma model
in which all four parameters can be exhibited.
The goal is to recover the claim that a singularity only arises
when all four parameters have special values and to learn
something about the singularity.

Most of the construction and analysis of the linear sigma model are quite
similar to the discussion of the $(2,2)$ case in \ref\otherwitten{E.
Witten, ``Phases Of $N=2$ Models In Two Dimensions,'' Nucl. Phys.
{\bf B403} (1993) 159.}, so we
will be brief.  The model we will discuss is a two-dimensional
$(4,4)$ globally supersymmetric theory consisting of
a $U(1)$ gauge theory coupled to two hypermultiplets $H_i$, $i=1,2$,
of the same charge.  From an $N=2$ point of view (in what follows
we count supersymmetries in two dimensions, so what we call
$N=2$ and $N=4$ are  related
to $N=1$ and $N=2$ in four dimensions), each $H_i$ consists
of a chiral multiplet $M^i$ of charge 1 and another chiral multiplet
$\tilde M_i$ of charge $-1$.  The $(4,4)$ $U(1)$ gauge multiplet
contains four scalars $\phi_i$ (as one can see by dimensional reduction
from more familiar facts in four or six dimensions).
The potential energy of the theory is
\eqn\unbo{ V={1\over 2e^2}\left(\vec{D}(H)-\vec {r}\right)^2 +{1\over 2}
|H|^2|\phi|^2.}
This formula is analogous to equation (3.2)  in \otherwitten\ for the
$(2,2)$ case.  The notation is as follows.  For $N=4$ in      two
dimensions, there are three $D$ functions $\vec{D}(H)$, quadratic and
homogeneous
in the components of $H$, generalizing the more familiar one $D$
function for two-dimensional $N=2$ theories.  (They transform as a vector
under an $SU(2)$ $R$ symmetry of the model that will be described later.)
The   three constants
$\vec{r}$ generalize the familiar Fayet-Iliopoulos interaction
of $N=2$ theories.  The four relevant operators associated with
the $A_1$ singularity are in fact the three components of $\vec{r}$ and
the $\theta$ angle of the $U(1)$ gauge theory.

The space of zero energy classical states
 with $H\not=0$ (and therefore $\phi=0$) is obtained
by setting $\vec{D}-\vec{r}=0$ and diving by the gauge group $U(1)$.
(The combined operation is the hyper-Kahler quotient
\ref\hitchin{N. J. Hitchin, A. Karlhede, U. Lindstrom, and M. Rocek,
``Hyperkahler Metrics And Supersymmetry,'' Comm. Math. Phys.
{\bf 108} (1987) 535.}, which was discovered in precisely the
present context.)  Let us carry this out explicitly
for the case that $\vec r = 0$.  From an $N=2$ point of view, the
three $D$ terms are the real and imaginary part of
 a holomorphic function of chiral superfields
\eqn\umbo{D_+=M^1\tilde M_1+M^2\tilde M_2}
and the usual $N=2$ $D$ term $D_0=\sum_i(|M^i|^2-|\tilde M_i|^2).$
Dividing by $U(1)$ and setting $D_0=0$ is equivalent (according
to geometric invariant theory \ref\mumford{D. Mumford and J. Fogarty,
{\it Geometric Invariant Theory} (Springer, 1982).})
to working with the $U(1)$ invariant holomorphic functions
$S^i{}_j=M^i\tilde M_j$.  Upon setting $D_+=0$,  there are
three such independent functions, $A=M^1\tilde M_1=-M^2\tilde M_2$,
$B=M^1\tilde M_2$, $C=M^2\tilde M_1$.  These obey the identity
\eqn\uhbo{A^2+BC = 0 .}
This complex equation in  ${\bf C}^3$ is a standard description of the
$A_1$ singularity, so we have established the fact that the classical
moduli space of $\phi=0$ vacua, at $\vec r=0$,  is $\R^4/{\bf Z}_2$.
If one repeats the computation at $\vec r\not= 0$, one gets a
non-singular space, exhibiting the $\vec r$ as the three parameters
associated with deforming and resolving the singularity. How the theta
angle enters the story will be seen momentarily.

So far we have discussed only the zero energy states of $\phi=0$.
What about zero energy states of $\phi\not= 0$?  Inspection of \unbo\
shows that such states exist only if $H=0$, and therefore that one
needs also $\vec{r}=0$.  Classically, these are sufficient requirements,
but quantum mechanically, as explained in \otherwitten, one requires
also $\theta=0$.  The reason for this is that on the branch of $\phi\not=
0$ but $H=0$, the low energy theory is a free $U(1)$ gauge theory;
turning on a non-zero theta angle gives a term in the energy
$|\theta/2\pi|^2$.  So the Coulomb branch of zero energy
states with $\phi\not= 0$, $H=0$\foot{I will use
the terms ``Coulomb branch'' and ``Higgs branch'' that were applied
in \wstwo\ to related theories in four dimensions, but the meaning
is rather different in two dimensions: because of two-dimensional
infrared divergences, these branches do not
parametrize a family of quantum vacua; rather, they are target
spaces of low energy supersymmetric sigma models.}
  exists only for $\vec{r}=\theta=0$.

Now the Higgs branch -- that is, the branch of low energy states
with $H\not= 0$ -- presumably
flows for any values of $\vec{r},\theta$ to a  $(4,4)$
conformal field theory in the infrared.  Our question is:
for what values of $\vec{r},\,\theta$ is this conformal field
theory singular?  As explained in \otherwitten, a singularity can
only arise when the vacuum state on the Higgs branch can spread
onto the Coulomb branch.  The situation is most easily described if
(as in \otherwitten) we  work on a compact $\K3 $ manifold
that is developing an $A_1$ singularity rather than, as above,
working simply on $\R^4/{\bf Z}_2$.  (Unfortunately, working on
a compact $\K3$ would have made it difficult
to  explicitly exhibit the four parameters associated
with the singularity.)  Then we would simply say
that unless $\vec{r}=\theta=0$, the theory has a normalizable vacuum
state, which ceases to be normalizable at $\vec{r}=\theta=0$ when
the vacuum can spread onto the Coulomb branch.  On ${\bf R}^4/{\bf Z}_2$,
the vacuum is not normalizable to begin with, but the new
non-compactness from the appearance of the Coulomb branch still gives
a singularity.

So we have learned that a singularity appears in the conformal field
theory precisely upon setting all four parameters to zero -- and thus
the conformal field theory is singular precisely where, according
to \newaspinwall, the extended gauge symmetry appears.
We would like to learn more about the nature of the singularity.

To do so, as in the $(0,4)$ problem that was discussed above,
we want to look at the possible global symmetries that can appear
in the $(4,4)$ superconformal algebra in the infrared.  These
symmetries are very conveniently seen by starting in {\it six}
dimensions with a $U(1)$ gauge multiplet coupled to the two
hypermultiplets $H^i$.  There is a global $SU(2)$ symmetry $G$:
 the group of linear transformations of the        eight real
components of the $H^i$ that preserves the hyper-Kahler structure
and commutes with the gauge group.  The fact that $G$ preserves
the  hyper-Kahler structure means that it commutes with all the
supersymmetries and so will not be seen as an $R$ symmetry under
any conditions.  There
is also, already in six dimensions, an $SU(2)$  $R$ symmetry $K$;
it acts trivially on the gauge   fields (and non-trivially, therefore,
on their fermionic partners),   while the bosonic part of $H^i$
transforms with $K=1/2$.  Dimensional reduction from six to two
dimensions produces an extra   $SO(4)$ symmetry, which we  write
as $L\times L'$, with $L$ and $L'$ being copies of $SU(2)$.  $L$ and $L'$
act trivially on the bosons in $H^i$ (because they are scalars
in six dimensions), but the scalars in the two-dimensional
vector multiplet (because of the way they arise from a vector
in six dimensions) transform in the $(1/2,1/2)$ representation of
$L\times L'$.

Now we can analyze the possible $R$ symmetries.  A $(4,4)$
superconformal field theory will have  left and right-moving
 $SU(2)$ $R$ symmetries.   For a     conformal field
theory arising from the Higgs branch, these $R$-symmetries must
act trivially on the     scalar components of $H^i$.  From the
above description, the symmetries with the right properties are
$L$ and $L'$, and it is easy to see in perturbation theory
(valid for large $H$) that $L$ and $L'$ do emerge as the $R$ symmetries
on the conformal field theory of the Higgs branch.

Setting $\vec r=\theta=0$, we can also analyze the singularities
of the Coulomb branch.
 On the Coulomb branch, the $R$ symmetries must act trivially
on the     scalars in the vector multiplet, so $L$ and $L'$ are
forbidden; the only possibility from what we have seen above is
$K$.  Perhaps $K$ decomposes in the infrared into separately conserved
left and right-moving pieces.

Just as in our discussion of the $(0,4)$ case, the fact that
 different $R$ symmetries enter
in the superconformal algebra on the different branches
must mean that by the time one
flows to a conformal field theory, the  branches  no longer
meet as they do classically.  The most natural way for this to happen
is once again that in the conformal field theory limit,
the point $H=\phi=0$ should be infinitely far away from the rest
of the Higgs branch.

So we are led to look for a $(4,4)$ superconformal field theory,
with $\hat c=4$, and the following characteristics.  The model
should be a sigma model with a four-dimensional target space that
is asymptotic to $\R^4/{\bf Z}_2$ at spatial infinity, while also
one point has been
deleted from $\R^4/{\bf Z}_2$ and in some way projected to infinity.
Happily, such a conformal field theory is known.  It is essentially  the
so-called symmetric five-brane \harvey, which can be described
as a four-dimensional sigma model with a target space metric
that coincides with the $\rho=0$ limit of
 equation \guggle.
This (or more exactly its quotient by ${\bf Z}_2$ to get the right
asymptotic behavior)
has just the properties that we want.  (There is a puzzle,
however, about the presence in the symmetric five-brane of a $B$
field with non-zero field strength - absent for conventional $\K3$'s
at least in sigma model perturbation theory.  Perhaps it is a novel
non-perturbative effect.)

Our proposal, then, is that at a point of extended gauge symmetry,
the  singular behavior of the conformal field theory is that
it develops an infinite tube like that of the
symmetric five-brane.   Hopefully, this understanding of the
singularity may lead in future to a better understanding of how extended
gauge symmetry comes about.
One simple remark that can be made right away is that, no matter
how small the string coupling constant may be on most of the $\K3$,
it blows up (because of the linear
dilaton) as one goes down the infinite tube of the five-brane.
Thus, once the $\K3$ in conformal field theory
develops such a tube,  there is no further surprise
in the fact that -- no matter how small the string coupling constant
is -- there are quantum effects that do not get turned off.
The place where these  effects occur
 just moves ``down the tube'' as the string coupling
constant is made smaller.  This seems to shed some light on some of the
puzzles of the last few months.

\bigskip\noindent{\it Zero Area?}

Finally, I want to resolve a small paradox that this discussion may
present.
Classically, as one takes $\vec {r}\to 0$, a two-sphere collapses
to zero size -- and in the  ``two-brane'' picture,    this is why
massless charged gauge bosons appear.  At first sight we have lost
this explanation upon replacing ${\bf R}^4/{\bf Z}_2$ with
the symmetric five-brane.
But  what is written in  \guggle\ is (at $\rho=0$)
 the {\it sigma-model metric}
of the symmetric five-brane.  This metric is conformally flat,
as is evident in the way it has been written, and in fact the
{\it Einstein metric} of the symmetric five-brane is simply the
flat metric on ${\bf R}^4/{\bf Z}_2$.
So the ``collapsing two-sphere'' mechanism survives the better
understanding of the singularity -- but must be implemented in the
Einstein metric.

\subsec{Conifolds In Calabi-Yau Threefolds}

Now we will -- more briefly -- discuss conifolds in Calabi-Yau
threefolds in a similar fashion.   The general
argument really  applies  to any  isolated singularities that will
arise by varying the complex structure of a Calabi-Yau manifold.
Via mirror symmetry (or more explicitly via linear
sigma models \nref\silver{E. Silverstein and E. Witten, ``Criteria
For Conformal Invariance Of $(0,2)$ Models,'' Nucl. Phys.
{\bf B444} (1995) 161.}
\refs{\otherwitten,\silver}), a similar
story can be told for singularities that arise upon varying Kahler
parameters.

We consider a $(2,2)$ model in  two dimensions with chiral superfields
$P,X,Y,Z,$ and $T$ and superpotential
\eqn\idmo{W=P(XY+ZT-\epsilon).}
For $\epsilon=0$, the classical states of zero energy -- which
are precisely the critical points of $W$ -- are described by
\eqn\jidmo{XY+ZT=\epsilon,~~~~ P=0.}
For $\epsilon\not= 0$, the equation $XY+ZT=\epsilon$ describes
a smooth hypersurface $V_\epsilon$ in ${\bf C}^4$.  For $\epsilon=0$,
$V_\epsilon$ develops a conifold singularity.
The singularity does not in itself
show that the low energy
conformal field theory is singular; we are familiar
with classical singularities (such as orbifold singularities
at $\theta\not=0$) that do not correspond to singularities in conformal
field theory.  What really shows that a singularity appears
in the field theory is that precisely at $\epsilon=0$, a second
branch of critical points appears, with $P\not=0, \,\,X=Y=Z=T=0$.
We will call this the $P$ branch.  The vacuum constructed
on the original ``$V$ branch'' spreads onto the $P$ branch
and (even when such a conifold singularity  is embedded in
a compact Calabi-Yau manifold) its normalizability is lost.
In \silver, a pole in Yukawa couplings at $\epsilon=0$ was deduced
directly from the appearance of the $P$ branch.

Now let us use the $R$ symmetries to learn a little more about
the superconformal field theories to which these theories presumably
flow in the infrared.  To get the necessary $R$ symmetry,
\foot{We need separate left and right-moving $R$ symmetries for
a $(2,2)$ model in two dimensions.  One combination of these
(which if one constructs the model by dimensional reduction
from four dimensions arises
as the rotation of the two extra dimensions) is
present in all models of this kind, so there is precisely one
model-dependent symmetry to be described.}
we need
a holomorphic $U(1)$ action on $P,X,Y,Z,T$ under which $W$ has
charge two.  The only appropriate symmetry, for $\epsilon\not=0$, is the
one  that assigns charge two to $P$ and charge zero to $X,Y,Z,$ and $T$.
This must therefore be the $R$ symmetry for non-zero $\epsilon$, and
by continuity it will therefore  be the $R$ symmetry on the $V $
branch also at $\epsilon=0$. (That $W=0$ for the bosonic fields
in $X,Y,Z,$ and $T$ makes it possible for $W$ to be the $R$-symmetry
on this branch.)  Since this symmetry acts non-trivially
on the bosonic part of $P$, it must be that at $\epsilon=0$, by the
time one flows to a conformal field theory, the $P$ branch is disconnected
from the $V$ branch.  Precisely at $\epsilon=0$, the theory
has a new $R$ symmetry -- the one under which $P$ is neutral and the
other fields all have charge one -- which has the right properties
to appear in the superconformal algebra on the $P$ branch.

So we learn again -- as in the earlier discussion of $(0,4)$ and $(4,4)$
models -- that when one flows to conformal field theory, the various
branches are disconnected.  As before, the most plausible interpretation
of this is that the sigma model metric of the conformal field
theory at $\epsilon=0$ is an incomplete metric, with $X=Y=Z=T=0$
being infinitely far away.

\bigskip\noindent{\it Relation To Quantum Description}

In each  case that we have examined,  the target space apparently
 becomes effectively non-compact at the  point
where the conformal field theory is singular.  For instance,
in compactification on $\R^4\times V$, with $V$ a Calabi-Yau
three-fold, space-time is four-dimensional macroscopically
as long as $V$ is smooth and compact.  But when $V $ acquires
a conifold (or other) singularity, we have argued that the
sigma model metric on $V$ becomes incomplete, strongly indicating
that the space-time becomes at least five-dimensional macroscopically,
in the sigma model description.  It may be quite different
in the Einstein metric.
For the $(0,4)$ and $(4,4)$ cases, where the sigma model metric
at the point analogous to $\epsilon=0$ is known,
 precisely one new macroscopic dimension appears
in the sigma model metric, but not in the Einstein metric.
For threefolds, we have much less information.

The singularities we have found for $\K3$'s and Calabi-Yau
threefolds in this worldsheet treatment are much more drastic
than what has been argued quantum mechanically.  Let $\lambda$ be
the string coupling constant and $\epsilon$ a parameter
measuring the distance in coupling constant space from the singularity.
For $\epsilon=0$ with $\lambda\not=0$, it has been argued that
what happens at the singularity is that finitely many massless
particles appear (charged gauge bosons or charged hypermultiplets
for Calabi-Yau twofolds or threefolds).  For $\epsilon=\lambda=0$,
we are instead finding a noncompactness which means that infinitely
many particles are going to zero mass in the low energy description.
Turning on quantum mechanics makes the behavior much gentler; in
particular, the effective dimension of space-time is not changed.
Perhaps a better understanding of the singular behavior of the
conformal field theory would enable one to understand in a more
{\it a priori} fashion what happens quantum mechanically.

There are other reasons, apart from what I have given here,
for suspecting that infinitely many particles become massless
at $\lambda=\epsilon=0$.  First of all, {\it some} particles
must become massless in the conformal field theory when one
sets $\epsilon=0$, because for instance the one loop conformal
field theory calculation in \ref\vafa{C. Vafa, ``A Stringy Test
Of The Fate Of The Conifold,'' hepth-9505023.} develops a
singularity at $\epsilon=0$.  This singularity somehow comes
from massless elementary string states running around the loop
(and not from charged Ramond-Ramond black holes, which are not
present in the conformal field theory!).  The charged black holes
are probably the only natural way to get this sort of singularity
from the contributions of finitely many particles, so it is not
too surprising that the conformal field theory method of generating
this singularity would turn out to involve infinitely many
light states.

\newsec{Dimensional Reduction Below Four Dimensions}

The final subject that I will discuss here concerns an attempt
to apply some of the new string theory ideas
 directly to nature.
Recently, I suggested \ref\cosmo{E. Witten, ``Strong Coupling
And The Cosmological Constant,'' hepth-9506101, submitted to
Mod. Phys. Lett. A.} that the vanishing of the cosmological
constant in nature results  from the existence of an
interpretation of the four-dimensional world as a strong coupling
limit of a supersymmetric world in three dimensions.  The idea
is that a mode which a three-dimensional observer interprets
as the dilaton is interpreted by a four-dimensional observer
as the radius of the fourth dimension.  Thus in the strong coupling
limit of the three-dimensional theory, the world becomes four-dimensional
and the dilaton is reinterpreted as part of the four-dimensional
metric tensor, so that there is no dilaton in the four-dimensional
sense.  In three dimensions, for generic coupling, the cosmological
constant vanishes but \ref\onepage{E. Witten,
``Is Supersymmetry Really Broken?'' Int. J. Mod. Phys. {\bf A10} (1995)
1247.}
 the bosons and fermions are not degenerate;
the limiting four-dimensional world hopefully inherits these
properties.

A crucial question about this scenario is what the dynamics looks
like from the three-dimensional point of view, as one approaches
the limit of four dimensions.  In particular, one want to retain
the vanishing of the cosmological constant but very few other
implications of three-dimensional supersymmetry.  It is not clear
precisely how this can work.  I will here discuss instead a more
straightforward question, which is what things look like from
the {\it four}-dimensional point of view when one is near the
four-dimensional limit.  That is, we will consider the dimensional
reduction of the real world on ${\bf R}^3\times {\bf S}^1$, and
ask what one sees when the radius $R$ of the $\S^1$ is extremely large.

In doing so, we will assume that on ${\bf R}^4$, the only exactly
massless bosons are the photon and the graviton; one could extend
the discussion if one knew what additional massless bosons to consider.
It would similarly be somewhat natural to assume that the massless
fermions are a subset of the known neutrinos, though this is of
course far from certain.

Upon reduction on $\R^3\times \S^1$, the photon becomes a scalar
$\phi$  and a vector $a$.
The graviton similarly  decomposes  as a scalar $r$ (the fluctuating
radius of the $\S^1$), a vector $b$, and a three-dimensional graviton
which does not have have any
propagating modes.  The intention here is to discuss in turn the
dynamics of the modes $\phi$, $a$, $b$, and $r$ -- taking them
roughly in increasing order of subtlety.

(1) $\phi$ is really an angular variable, with $0\leq \phi\leq 2\pi$,
since it is best interpreted as the holonomy of the photon around
$\S^1$.  At the classical level, the energy of the vacuum is independent
of $\phi$.  Quantum mechanically, as electrons
are the lightest charged particles, the main influence of $\phi$
is on the vacuum energy of the filled Dirac sea of electrons.
This energy is minimized at $\phi=\pi$ \ref\hoso{Y. Hosotani,
``Dynamical Gauge Symmetry Breaking As The Casimir Effect,'' Phys.
Lett. {\bf 129B} (1983) 193.} so
$\phi$ will acquire that vacuum expectation value.  Expanding around
the minimum, the mass of $\phi$ is roughly of order $e^{-\pi mR}$
with $m$ the electron mass.

(2) The three-dimensional photon $a$ is massless in perturbation
theory.  If, however, as is generally believed, magnetic monopoles
exist in nature, then by thinking of the $\S^1$ direction
as ``time,'' a time-independent magnetic monopole on $\R^3\times \S^1$
is a localized object that can be interpreted as a kind of
instanton.  Three-dimensional $U(1)$ gauge theory with such
instantons (``compact QED'') was first studied by Polyakov
\ref\polyakov{A. M. Polyakov, ``Compact Gauge Fields And The
Infrared Catastrophe,'' Phys. Lett. {\bf 59B} (1975) 82.}
and has the remarkable property that the photon acquires a mass
-- a phenomenon most conveniently described in terms of a scalar
$u$ dual to $a$.  The mass of $u$ is roughly of order
$\exp(-\pi MR)$ with $M$ the mass of the lightest magnetic monopole
in nature.  If this phenomenon does occur, as one would expect,
then electric charges are subject to not just logarithmic but
linear confinement.

(3) Now we come to the second photon $b$ of the three-dimensional
world.  Though the  physics involved is not well understood,
it is very plausible that also in this case suitable instantons
exist and the $b$ field gets a mass.  In this case the charge that would
be subject to linear confinement is the one that comes from
rotations of the $\S^1$; the modes carrying momentum in the fourth
dimension would be confined!

(4) Finally, we come to the scalar $r$ that measures fluctuations in
the radius of the $\S^1$.  If the cosmological
constant vanishes in four dimensions, then
 the potential $V(r)$ for this
scalar vanishes for $r\to\infty.$   Corrections vanishing as a power
of $r$ for $r\to \infty$ can be computed
systematically by evaluating Feynman diagrams involving massless
particles only.  The leading correction for $r\to\infty$, for instance,
is the one-loop Casimir effect of the massless bosons and fermions
in nature, and is a multiple of
\eqn\plop{-{n_B-n_F\over r^3}}
with $n_B$ and $n_F$ the number of exactly massless bose and fermi
helicity states in nature.  Feyman diagrams of massless particles
with two or more loops give corrections of higher order in $1/r$.
To compute to order  $1/r^{3+n}$ needs to know the effective
Lagrangian of the exactly massless particles in nature including
all terms up to dimension $4+n$.  Thus, the more perfectly the
low energy effective action of nature is known, the more precisely
one could work out the expansion of $V(r)$ in powers of $1/r$.

Without any further theory,
 one would assume that $V(r)$ is non-zero except
in the limit of $r\to \infty$; we know from experimental bounds
on the cosmological constant that $V(\infty)$ is zero or at least
incredibly small.  The scenario
in \cosmo, however, at least in the form presented there, implies
that $V(r)$ is identically zero.  This is indeed the four-dimensional
analog of the three-dimensional statement that because of unbroken
supersymmetry the vacuum energy is zero for any value of the dilaton
field.  Thus, this scenario makes the remarkable prediction that
the vanishing of the cosmological constant is only the first
of an infinite series of vanishing phenomena that might mystify a
low energy observer.  The second prediction is that $n_B-n_F=0$,
and subsequent predictions, involving the $r$ dependence of Feynman
diagrams with two or more loops, could be worked out given sufficient
knowledge of the low energy world.  This framework, then, certainly
has some predictive power, if not too much.

%The lectures by Sen and Schwarz at {\it Strings `95} actually
%made me wonder about another way that duality symmetries could
%be relevant to dimensional reduction below four dimensions.

\bigskip

I would like to thank M. Dine, D. J. Gross,
D. Morrison, M. Peskin, M. Reid,
N. Seiberg, S. Shenker,
E. Silverstein, and A. Strominger
for discussions concerning some of these matters.

\listrefs

\end